\documentclass[preprint]{aastex}
\usepackage{booktabs}

\begin{document}

\title{Spatio-temporal Characteristics of Very Long-periodic Pulsations in Solar Metrewave Bursts: Implications for their Origins}

\author{Dong~Li$^{1,2}$, Lei~Lu$^{1}$, Jingye~Yan$^{2}$, Xinhua~Zhao$^{2}$, Bing~Wang$^{3}$, Chengming~Tan$^{2}$, Jianping~Li$^{1}$, Zongjun~Ning$^{1}$}
\affil{$^1$Purple Mountain Observatory, Chinese Academy of Sciences, Nanjing 210023, China \\
       $^2$State Key Laboratory of Solar Activity and Space Weather, National Space Science Center, Chinese Academy of Sciences, Beijing 100190, China \\
       $^3$Laboratory for Electromagnetic Detection, Institute of Space Sciences, Shandong University, Weihai, 264209, China}
     \altaffiltext{}{Correspondence should be sent to: lidong@pmo.ac.cn}
\begin{abstract}
We traced the origin of very long-periodic pulsations (VLPs) in
type-I burst chains on 2024 February 14. Seven successive and
repetitive pulsation structures appeared in radio dynamic spectra in
the metric waveband, which were simultaneously measured by CBSm,
DART, and MUSER-L. A quasi-period at about 160$^{+11}_{-6}$~s,
determined by the fast Fourier transform, was detected in the
frequency range of about 210-280~MHz. Imaging observations from DART
and SDO reveal that the type-I burst chains occur above two groups
of sunspot umbrae connected by coronal loops. A quasi-period of
approximately 170~s was also identified in the sunspot umbrae and
coronal loops. The burst chains exhibit strong circular polarization
and high brightness temperature, and they show spatiotemporal
correlation with emerging magnetic flux. The number
densities at the loop top and double footpoints can produce radio
emission and generate type-I burst chains in the frequency range of
210-280~MHz. Our observations support the scenario that plasma
emission serves as the primary generation mechanism of type-I
bursts, with VLPs most likely being modulated by the slow
magnetoacoustic waves originating from sunspot umbrae. The observed
frequency drift of burst chains may reflect the density attenuation
along coronal loops.
\end{abstract}


\section{Introduction}
Solar radio bursts, among the earliest phenomena targeted in radio
astronomy, are classified into five types based on their frequency
characteristics \citep{Wild63}. Types-II, III, and IV bursts are
particularly relevant for space weather studies, as they are always
associated with major solar eruptions, i.e., solar flares and
coronal mass ejections \citep[CMEs;
e.g.,][]{Lu22a,Lu22b,Azzollini25,Kumari25}. Type-V bursts are
relatively rare and can be difficult to identify, especially when
there are some other bursts present, e.g., emerging from type-III
bursts \citep{White24}. In contrast, type-I bursts are typically
associated with active regions or sunspots, and they are usually
independent of solar flares \citep{Mercier84,Li17}. Evidence also
suggests that type-I bursts may be modulated or generated by CMEs
\citep{Iwai12}. Type-I bursts are always detected in the metric
waveband, consisting of a burst component superposed on a continuum
component. The continuum component, also referred to as a noise
storm, often appears at frequencies below a few hundred MHz and
persists for several hours \citep{Dez11,Suresh17}. This prolonged
duration indicates that the continuum component is likely associated
with energetic electrons trapped in closed magnetic field lines at
the coronal height \citep{Melrose80}. Conversely, the burst
component of type-I bursts is characterized by sub-second durations
and extremely narrow wavebands, typically appearing in drifting
chains and covering a frequency range of about 10--20~MHz
\citep{Mondal24,Yang25}. Type-I burst radiation typically exhibits
strong circular polarization \citep{Mugundhan18,McCauley19}, which
may result from plasma emission from nonthermal electrons
accelerated during magnetic reconnection~$¡ª-$~specifically
involving, emerging magnetic flux and moving magnetic fields
\citep{Spicer82,Zwaan85,Li17}. Type-I bursts are intriguing because
their continuum components persist in the corona over long
timescales; however, the lack of clear association with energy
releases at other wavelengths makes their physical nature very
difficult to understand \citep{Elgaroy76,White24}.

Solar quasi-periodic pulsations (QPPs) frequently manifest as
a number of successive and repeated pulsations in time-dependent
fluxes \citep{Zimovets21}. They have been observed across a wide
range of wavelengths, including radio, white light, ultraviolet (UV),
extreme ultraviolet (EUV), soft/hard X-rays (SXR/HXR), and
$\gamma$-rays
\citep[e.g.,][]{Nakariakov18,Li22,Kim25,Li24,Lis25,Purkhart25,Song25,Ashfield26}.
QPPs can be detected not only in intensity curves of solar flares
\citep{Shi24,Li25,Zimovets25} and bright points \citep{Lid22,Lim25},
but also in the time series of radio bursts \citep{Yang25}. QPPs
remain a topic of ongoing interest and have gained increasing
importance, as they offer insights into time characteristics and
plasma emission~$¡ª-$~including the interior structure and physical
conditions of the radiation source, as well as the properties of the
propagation media \citep{Tan08}. The observed QPPs are often
interpreted as magnetohydrodynamic (MHD) waves \citep{Nakariakov20},
they can also be triggered by the oscillatory magnetic reconnection
\citep{Li25,Li26,Schiavo25}. MHD waves may directly drive fluctuations in
local densities and temperatures \citep{Yuan15,Nakariakov19}, or
modulate plasma emission via variations in the angle
between the line of sight and local magnetic fields
\citep{Kohutova20}. Oscillatory magnetic reconnection can
periodically accelerate energetic electrons, leading to
quasi-periodic precipitation of these particles
\citep{Jakimiec10,Li25a}.

The quasi-period of QPPs, which is generally a varying instantaneous
period, is observed to be distributed across a wide time
range~$¡ª-$~from several tens of milliseconds to a few hundred
seconds \citep{Nakariakov18,Yu19,Collier24,Li25b,Yang25,Karlicky26}.
According to the timescale of quasi-periods (P), \cite{Tan10}
proposed a classification: (1) very long-periodic pulsations (VLPs,
P$>$100~s), (2) long-periodic pulsations (LPPs, 10$<$P$<$100~s), (3)
short-periodic pulsations (SPPs, 1$<$P$<$10~s), (4) very
short-periodic pulsations (VSPs, P$<$1~s). The QPPs with various
quasi-periods are frequently detected in solar radio emission,
particularly in the microwave range of decimeter and centimeter
\citep{Tan10}. However, QPPs are rarely reported in the metric
waveband, especially for the VLPs event. This is attributed to the
limitations of observational instruments, namely the lack of
high-sensitive radio telescopes in the metric waveband. In this
Letter, we investigate VLPs during solar burst chains in the metric
waveband, unraveling their origins.

\section{Observations}
We analyzed solar metrecwave burst chains on 2024 February 14. The
burst chains were simultaneously measured by the Chashan Broadband
Solar radio spectrometer at meter wavebands
\citep[CBSm;][]{Chang24}, Mingantu Ultrawide SpEctral
Radioheliograph at low frequencies \citep[MUSER-L;][]{Yan24} and the
DAocheng Radio Telescope \citep[DART,formerly DSRT][]{Yan23}. We
also employed observations from space-based instruments: the Macau
Science Satellite-1 \citep[MSS-1;][]{Zhang23}, the Geostationary
Operational Environmental Satellite (GOES), the Hard X-ray Imager
\citep[HXI;][]{Su24} aboard the Advanced Space-based Solar
Observatory (ASO-S), the Atmospheric Imaging Assembly
\citep[AIA;][]{Lemen12} and the Helioseismic and Magnetic Imager
\citep[HMI;][]{Schou12} Instrument on the Solar Dynamics Observatory
(SDO). This comprehensive dataset from both ground-based and
space-based instruments enabled us to study the temporal, spectral,
and spatial characteristics of the metrecwave burst chains and their
association with solar activities. A comprehensive summary of all
instruments and their key parameters is presented in
Table~\ref{tab1}.

\begin{table*}[ht]
\addtolength{\tabcolsep}{6pt}
\centering \caption{Instruments and parameters employed in this study.}
\label{tab1}
\begin{tabular}{cccccc}
\toprule
Instrument &  Waveband     &  Cadence   &  Description      \\
\midrule
CBSm       &  90-600~MHz    &   $\sim$0.1~s &  Radio spectrum     \\
\midrule
MUSER-L    &  30-400~MHz    &   $\sim$0.1~s &  Radio spectrum     \\
\midrule
DART       &  140-460~MHz    &   $\sim$0.4~s &  Radio spectrum     \\
DART       &  205/223/238/285~MHz  & $\sim$1~s &  Radio image    \\
\midrule
MSS-1      &  0.5-8~{\AA}   &   1~s      &  SXR spectrum     \\
\midrule
           &  1-8~{\AA}     &   60~s     &  SXR flux          \\
GOES       &  0.5-4~{\AA}   &   60~s     &  SXR flux          \\
\midrule
SDO/AIA    &  171/131{\AA}  &   12~s     &  EUV image         \\
SDO/AIA    &  1700{\AA}     &   24~s     &  UV image          \\
\midrule
SDO/HMI    &  IC            &   45~s     &  Continuum image    \\
SDO/HMI    &  M             &   45~s     &  LOS magnetogram     \\
\midrule
ASO-S/HXI  &  10-30~keV     &   1~s      &  HXR flux/image          \\
\bottomrule
\end{tabular}
\end{table*}

\section{Results}
Figure~\ref{over}~(a) shows the SXR/HXR fluxes recorded by GOES,
MSS-1 and ASO-S/HXI from 06:45~UT to 07:55~UT. The SXR light curves
exhibit two peaks at about 06:54~UT and 07:35~UT on 2024 February
14, indicating two solar flares. The largest peak in the HXR flux,
as observed by ASO-S/HXI, is attributed to its passages through the
South Atlantic Anomaly (SAA). The GOES~1$-$8~{\AA} flux indicates
that the flares are classified as C3.6 and M1.0, with no obvious
eruptions observed in the SXR/HXR waveband between them. Conversely,
clear radio bursts in metric wavelengths are observed between the
C3.6 and M1.0 flares. The two solar flares occurred in the active
region NOAA 13576 in the southern hemisphere, as shown in
Figure~\ref{add1}. Figure~\ref{over}~(b) and (c) present the radio
dynamic spectra in metric wavelengths measured by CBSm and MUSER-L
from 07:00 to 07:20 UT. Both dynamic spectra display numerous solar
radio bursts in the frequency range of about 200--300 MHz.
Particularly, about seven groups of metrecwave bursts appear between
about 07:03~UT and 07:18~UT, coinciding with the time interval
between the C3.6 and M1.0 flares. The metrecwave bursts mostly drift
from the higher to lower frequency, while the bursts~2 show a
reverse drift. The average drifting rates are estimated in the range
of 0.7-1.5~MHz~s$^{-1}$. These metrecwave bursts exhibit similar
characteristics, such as a narrow bandwidth (10-30~MHz) and a short
timescale (10-30~s). We note that a long gap appears between
metrecwave bursts 2 and 3. This may because that only burst~2 drifts
clearly from lower to higher frequencies. The metrecwave bursts are
characterized by a large number of spiking structures, indicating
that they are mostly generated by plasma emission. Panel (d) shows
the dynamic spectrum of the polarization measured by DART. It is
evident that these metrecwave bursts exhibit a strong circular
polarization. All those dynamic spectra suggest that each group of
metrecwave bursts may consist of numerous type-I bursts, forming
type-I burst chains rather than an individual type-I burst.

To determine the source morphology of the type-I burst chains, we
present multi-wavelength snapshots from SDO/AIA, SDO/HMI, and DART,
as shown in Figure~\ref{img}. Panels~(a1)-(a3) show sub-maps with a
field of view (FOV) of about 800\arcsec~$\times$~800\arcsec~in
wavelengths of HMI continuum intensity (IC), AIA~1700~{\AA} and
171~{\AA} during the burst chain~2. Overlaid color contours
represent metric radiation measured by DART in different
frequencies, with contour levels set at 50\% of the local maximum.
The radiation source at DART 238 MHz is located above two groups of
sunspot umbrae (hot pink contours) connected by coronal loops, as
outlined by the cyan rectangle. We note that the morphology of
metric sources remains stable during the lifetime of burst chain~2
at 238 MHz, as shown by the color contours in panel (a1). That is,
the source morphology shows minimal temporal variation throughout
the event. The radiation sources at DART~238~MHz and 223~MHz nearly
overlap at the same time, but they slightly deviate from those at
205~MHz and 285~MHz, indicating minor morphological differences
across frequencies. This can be explained by the narrow bandwidth of
burst chain~2, which is primarily enhanced between about 220~MHz and
250~MHz. Figure~\ref{img}~(b1)-(b3) and (c1)-(c2) show sub-maps with
the same FOV in HMI IC and line-of-sight (LOS) magnetogram (M),
AIA~131~{\AA} and 171~{\AA} wavelengths during the burst chains~4
and 6. They exhibit similar characteristics to the burst chain~2,
for instance, the source morphology is stable in the same frequency
across different times, but it is a slight deviation at various
frequencies, indicating a shared source region and narrow bandwidth
during the burst chains. The animation of anim.mp4 further confirms that
all seven burst chains share the same source region, and the
morphology of the metrewave burst chains is nearly identical.
Moreover, these burst chains are located above the same sunspots
connected by a series of coronal loops (S1), which are rooted in
positive and negative magnetic fields, respectively. It can be
concluded that all these burst chains are located above two groups
of sunspot umbrae in the northern hemisphere, which is far from the
flaring active region (Figure~\ref{add1}). Panel~(d) gives the
magnetic configurations of the active region that produces the
metric radiation, derived from a potential field source surface
(PFSS) extrapolation. The active region is primarily filled with
closed magnetic field lines rooted in positive and negative
polarities, with the metric sources linked by some magnetic field
lines, as indicated by the magenta arrow. Given a semi-circular
profile of coronal loops \citep{Tian16}, the loop length ($L$) is
estimated to be about 130~Mm.

Figure~\ref{flux}~(a) and (b) show the time series in multiple
wavelengths measured by DART, CBSm, HMI, and AIA. The time series of
bright temperature (BT) and circular polarization is extracted from
the metric source region (outlined white rectangle in
Figure~\ref{img}) in the frequency of DART~238~MHz, and the
opposite sign of the polarization degree indicates dominance of the
left-handed polarized component of the radio flux. All seven burst
chains cover the frequency range of 238~MHz, although some burst
chains(e.g., 3 and 7) exhibit relatively weak radiation, as
indicated by the white-dashed line in Figure~\ref{over}~(b). To
maximize coverage of the burst chains, the normalized flux
integrated over the frequency range of 210-280~MH is extracted from
the dynamic spectrum of CBSm. The resulting light curves from DART
and CBSm exhibit consistent trends. {Moreover, the time
series of brightness temperature from the DART instrument at 238 MHz
matches that of the polarization degree, indicating that the burst
chains exhibit circular polarization. All burst chains show a high
degree of circular polarization, reaching up to approximately 43\%.
The brightness temperature of these burst chains can exceed
10$^{11}$~K. These findings support the plasma emission model for
metrewave bursts \citep{Lu22a,Lu22b}. It is intriguing that the
metric flux of brightness temperature closely follows the variations
in net magnetic flux integrated over the active region (outlined by
cyan rectangle in Figure~\ref{img}) from the HMI LOS magnetogram.
Specifically,the pulsations of burst chains exhibit a
one-to-one correspondence with the pulses of net magnetic flux,
suggesting that metrewave bursts are highly associated with emerging
magnetic flux \citep{Zwaan85}. It should be pointed out that
peaks 3 and 7 appear to occur during periods of stable magnetic
flux. This may be because the amplitude of magnetic flux
fluctuations was too small for HMI to measure, given its low time
resolution of only 45~s. Panel (b) also plots the local UV/EUV
fluxes integrated over the active region in wavelengths of
AIA~1700~{\AA} and 131~{\AA}. The AIA~1700~{\AA} flux displays five
peaks during about 07:01-07:20~UT, but these do not align with the
seven metric pulsations. In contrast, the AIA~131~{\AA} flux remains
relatively gentle throughout the burst chains, with no clear peaks
corresponding to the metric pulsations. These observations suggest
that metrewave bursts are not triggered by small-scale brightenings
in EUV/UV wavebands. Figure~\ref{flux}(c) and (d) present the
detrended time-distance (TD) images after removing the 4-minute
running average along coronal loops (S1) in AIA~171~{\AA}. This
detrending, applied using a 4-minute window, highlights the
signature of slow waves with the quasi-period of about 3 minutes
\citep{Tian12,Meadowcroft24}. The start (S) and end (E) sub-parts of
the TD map near the double footpoints were selected because
slow-mode MHD waves are significantly stronger there than in the
loop-top region. The slow waves are identified as a series of
repeated and successive slanted stripes. In panels~(c) and (d),
these slanted stripes appear to travel in opposite directions,
indicating that the slow waves propagate along coronal loops. The
overplotted curves are derived from detrended images integrated over
two short lines on the left, which is used for further analysis.

In order to identify the quasi-period of VLPs during the type-I
burst chains, we applied the fast Fourier transform (FFT) to the raw
time series across multiple wavebands, as shown in
Figure~\ref{per}~(a)--(e). The dominant period was determined from
the peak Fourier power exceeding the 95\% confidence level, with
error bars defined by the intersection points between the Fourier
power and the confidence level near the peak. A quasi-period at
about 160$^{+11}_{-6}$~s (P1) is found in the metric emission of
CBSm~210$-$280~MHz and DART~238~MHz, while quasi-periods of
approximately 170$^{+16}_{-5}$~s and 166$^{+9}_{-5}$~s are detected
in the net magnetic flux of HMI LOS M and coronal loops of
AIA~171~{\AA}, respectively. Additionally, a quasi-period at about
275$^{+11}_{-10}$~s (P2) is measured in the AIA~1700~{\AA} flux. The
quasi-periods observed in the metric emission, net magnetic flux,
and coronal loops are quite similar, indicating that their
periodicity is modulated by the same process or shares a common
origin. Panel~(f) shows the spatial distribution of the normalized
Fourier power averaged over the spectral components between 150--210
s. Here, a Fourier transform is applied to each pixel of the HMI IC
data using the pixelised wavelet filtering (PWF) method
\citep{Sych08}. The spectral component at periods of 150--210~s
consistently appears in sunspot umbrae, as outlined by the magenta
contours. This suggests that the slow magnetoacoustic waves
originate from umbral oscillations at sunspots. Finally, the
quasi-period detected in AIA~1700~{\AA} is close to the 5-minute
oscillations observed in the quiet Sun, which may originate from the
solar P-mode waves in the photosphere.

Using an improved version of the sparse inversion code
\citep{Cheung15,Su18}, we performed the differential emission
measure (DEM) analysis, and the DEM(T) distribution for each pixel
was derived from six EUV channels of SDO/AIA sub-maps with a FOV of
about 800\arcsec~$\times$~800\arcsec, as shown in Figure~\ref{dem}.
The DEM solution provided rich information by mapping thermal
plasmas in a broad range of about 0.31--30 MK, and the DEM
uncertainty could be estimated from a Monte Carlo (MC) simulation,
i.e., the standard deviation of 100 MC realizations. Panels~(a) and
(b) draw the EM maps integrated from 0.5~MK to 2.8~MK within which
the coronal loop is normally detected, and the emission is
accumulated along the LOS into the observed intensity. The coronal
loop is clearly found in the temperature range of 0.5--1.8~MK, while
it cannot be seen in the temperature range of 1.8--2.8~MK. We
selected three positions with a small FOV of about
1.2$^{\prime\prime}$$\times$1.2$^{\prime\prime}$ to display DEM
profiles along the coronal loop, labeled as double footpoints (ft1
and ft2) and loop top (lp). Figure~\ref{dem}~(c) shows the DEM
profile as a function of plasma temperature, and the error bars
represent MC-derived uncertainties. The DEM profiles derived at
double footpoints exhibit three peaks (I, II, III) at about 0.7~MK,
1.4~MK, and 2.3 MK, and that at loop top display two peaks at about
1.4~MK, and 2.3~MK. Since the coronal loop is clearly seen in the
temperature range of 0.5--1.8~MK (panel~a), we therefore assume that
the high-temperature peak at about 2.3~MK is mainly originated from
the radiation of the diffuse background in AIA~211~{\AA}
\citep[cf.][]{Li20}. For a comparison, the DEM profile at a
reference position (bk) in the background corona is shown for
crossvalidation, as indicated by the tomato curve. It can be noted
that the background DEM profile indeed only reveals one prominent
peak (III) at about 2.3~MK, confirming that the coronal loop has a
low temperature range, i.e., 0.5--1.8~MK, as outlined by the pink
rectangle.

The emission measure (EM) can be calculated by integrating
the DEM(T) over temperatures (T), as shown in Eq.~\ref{eq1}. Here,
the temperature range between 0.5$-$1.8~MK is used for integrating,
because it is the effective temperature range of the coronal loop.
On the other hand, the EM could be regarded as the product of the
square of the number density and LOS depth ($d$). The LOS depth is
approximated with the loop width, referred as the full width at the
half maximum (FWHM) measured perpendicular to the loop axis, and the estimated loop width is about 15.6~Mm. Thus, the number density can be estimated along the coronal loop with
Eq.~\ref{eq2}, yielding $\sim$1.1$\times$10$^{9}$~cm$^{-3}$ and
$\sim$9.8$\times$10$^{8}$~cm$^{-3}$ at double footpoints  in the coronal height, and
$\sim$5.3$\times$10$^{8}$~cm$^{-3}$ at the loop top. This indicates
that the number density gradually decreases from double footpoints to the loop
top along the coronal loop, similar to the pervious finding along an
ultra-long loop \citep{Li20}. We note that the number densities at double footpoints and the loop top exhibit only a slight difference, indicating that the number density varies minimally with height along the coronal loop.

\begin{equation}
\centering
 EM = \int DEM(T) dT,
 \label{eq1}
\end{equation}

\begin{equation}
\centering
 n_e = \sqrt{EM/d},
\label{eq2}
\end{equation}
\noindent where, $n_e$ is the number density, $d$ represents the LOS depth.

\section{Conclusion and Discussion}
We systematically analyzed the quasi-periodicity of type-I burst
chains on 2024 February 14, which were simultaneously measured by
three radio telescopes in the metric waveband: CBSm, DART, and
MUSER-L. The metrewave bursts were primarily detected in the
frequency range of 210--280 MHz and appeared as drifting chains,
termed `burst chains'. These burst chains exhibited no
spatiotemporal connection with solar flares, that is, they occurred
at different active regions and at different times. Furthermore, the
burst chains showed strong circular polarization, with a degree of
polarization reaching up to $\sim$43\%. These observational features
are consistent with the characteristics of type-I bursts
\citep[e.g.,][]{Mercier84,Mugundhan18,McCauley19,White24}. However,
the timescale of an individual type-I burst is always shorter than
1~s, and its bandwidth is typically narrow, with a frequency scale
of several MHz \citep{Mondal24,Yang25}. In our observations, the
bandwidth falls within the frequency range of about 10--30 MHz, and
the time duration exceeds 10~s, which are inconsistent with the
typical values of an individual type-I burst. This suggests that the
metrewave bursts observed here may consist of a large number of
type-I bursts, forming what we refer to as `type-I burst chains'.
That is, each burst chain is actually a group of type-I bursts.
Notably, these burst chains cannot be regarded as drifting pulsating
structures (DPSs). DPSs are often identified as signatures of
plasmoids in decimetric and centimetric wavebands and are
characterized by a bandwidth of a few hundred MHz
\citep{Karlicky25}. Also, a majority of DPSs are found to be
associated with SXR peaks and HXR bursts, implying that DPSs may be
attributed to a process of fast energy release of solar eruptions
\citep{Nishizuka15}. The characteristics of DPSs are fundamentally
different from those of the burst chains observed in the metric
waveband, and these burst chains are unrelated to solar eruptions in
wavebands of SXR/HXR and EUV. The type-I burst chains consist of
seven successive and repeated metrewave bursts, which could be
regarded as QPPs. The long gap between metrewave bursts 2
and 3 implies the irregular nature of QPPs, suggesting the presence
of a characteristic timescale of the variations, termed as
`quasi-period' \citep[cf.][]{Nakariakov18}. By applying the FFT technology, a quasi-period at about 160$^{+11}_{-6}$~s is identified
in the frequency range of 210--280~MHz as measured by CBSm and DART.
According to the classification proposed by \cite{Tan10}, QPPs with
a quasi-period of 160~s can be classified as VLPs (i.e., P$>$100~s).
It is particularly intriguing to explore the generation mechanism of
type-I burst chains and the origin of VLPs. The bright temperature
integrated over the metric source area at 238~MHz reaches
10$^{11}$~K, and each burst chain consists of a large number of
spikes, suggesting that the radiation mechanism of type-I bursts is
plasma emission \citep{Lu22a,Lu22b}. The time series of type-I burst
chains matches with the net magnetic flux measured by the HMI LOS
magnetogram, showing a pulse-to-pulse correspondence.
Furthermore, the metric source area located above the active region
where the net magnetic flux is integrated. The slight variations in
the metric source area during type-I burst chains occur across the
observed frequencies, which may indicate that magnetic reconnection
takes place at different coronal heights and travels along coronal
loops \citep{Mondal24}. The type-I burst chains are temporally and
spatially correlated with emerging magnetic flux \citep{Spicer82},
demonstrating that these bursts likely originate from plasma
emission driven by the acceleration of energetic electrons due to
photospheric magnetic activities \citep{Zwaan85,Li17}. Additionally,
our study reveals an intriguing correlation between emerging
magnetic flux and type-I burst chains.

A key motivation of this study is to explore the origin of VLPs in
type-I burst chains. The quasi-period of VLPs is very close to the
3-minute oscillations observed in sunspot umbrae, and the type-I
burst chains are located above two groups of sunspot umbrae.
Therefore, we investigate the causal relationship between VLPs and
umbral oscillations, as well as their connection. Using the HMI IC
data, a quasi-period centered at about 180~s is observed in sunspot
umbrae, indicating the presence of slow magnetoacoustic waves in the
photosphere \citep{Sych08,Li24,Li25a}. Moreover, umbral oscillations
are detected in the temporal range of 06:30-07:30~UT, covering the
time interval before, during, and after the occurrence of type-I
burst chains. This observation suggests that VLPs may stem from the
slow magnetoacoustic waves
\citep{Yuan15,Nakariakov19,Wang21,Karlicky26}. Subsequently, two
groups of sunspot umbrae are found to be connected by coronal loops
in wavebands of AIA~171~{\AA} and 131~{\AA}. The PFSS extrapolation
further confirms that the metric sources are linked by closed
magnetic field lines, which are rooted in positive and negative
magnetic fields, respectively. Interestingly, a similar quasi-period
is measured at the double footpoints of coronal loops, implies that
slow magnetoacoustic waves propagate along the coronal loops
\citep{Meadowcroft24,Li25b}. Our observations provide sufficient
evidence that VLPs are modulated by slow magnetoacoustic waves,
which originate from umbral oscillations at sunspots.

Now, the question arises: how can VLPs be effectively modulated by
slow magnetoacoustic waves, given that the slow wave is a weak
signal, while VLPs are very clearly observed in type-I burst chains?
In the coronal environment, if the duration of heating pulsations
($\triangle t_{\rm H}$) is much smaller than the loop sound crossing
timescale ($\tau_{\rm s}$) \citep[cf.][]{Reale19}, then slow
magnetoacoustic waves can directly and easily modulate the obvious
or large-amplitude QPPs. In our case, the loop sound crossing
timescale is precisely constrained at about 2300~s (Eq.~\ref{slow1})
using coronal loops observed in AIA~171~{\AA}. Assuming $\triangle
t_{\rm H} \simeq P$ (where $P$ is the quasi-period), the observed
quasi-period of 160~s is indeed much less than the loop sound
crossing timescale. Thus, the VLPs observed in type-I burst chains
can be directly modulated by slow magnetoacoustic waves propagating
along coronal loops.
\begin{equation}
  \triangle t_{\rm H} \ll \tau_{\rm s} \approx 5~\frac{L({\rm Mm})}{\sqrt{0.1\cdot~T({\rm MK})}}.
  \label{slow1}
\end{equation}
\noindent Where, $L({\rm Mm})$ and $T({\rm MK})$ represent the loop
length and plasma temperature of coronal loops, with the units of
Megameter (Mm) and megakelvin (MK), respectively.

Finally, we discuss the mechanism of frequency drift in type-I burst
chains. The electron number density ($n_e$) in the metric source
region can be estimated using equation~\ref{ne1}:
\begin{equation}
\centering
n_{\rm e} = (\frac{f_{\rm p}}{8980})^{2}
\label{ne1}
\end{equation}
\noindent $f_{\rm p}$ is the radio frequency of local plasmas.
Type-I burst chains are clearly observed in the frequency range of
about 210--280~MHz, with corresponding electron number densities
estimated to about 5.5$\times$10$^8$---9.8$\times$10$^8$~cm$^{-3}$.
These densities agree with those measured at the loop top
and double footpoints connected by coronal loops, demonstrating
that the number density decreases progressively from double
footpoints to the loop top along coronal loops. Therefore, the frequency drift of radio burst chains is mostly due to the gradual decrease in plasma densities along the loop's height. Additionally, the frequency drift might arise from propagating disturbances that induce independent reconnections at
different heights, leading to the generation of localized nonthermal
electrons that gives rise to the observed type-I burst chains
\citep{Mondal24}. The changes in the source area of type-I burst
chains across observed frequencies suggest that magnetic
reconnection happens at multiple coronal heights. However, we cannot
identify fine-scale changes in the morphology of metric source,
largely due to the low-spatial resolution of DART images. Therefore,
it is challenging to conclude that the different reconnection
heights responsible for causing frequency drift. Nevertheless, the
frequency drift falls beyond the scope of this work, as our primary
focus is on exploring the origin of VLPs in type-I burst chains.

\acknowledgments The authors would like to thank the referee
for his/her inspiring comments. This work is funded by the National Key R\&D Program of China
2022YFF0503002 (2022YFF0503000), NSFC under grants 12573057 and
12250014, and the Strategic Priority Research Program of the Chinese Academy of
Sciences, Grant No. XDB0560000. The work is especially supported by the Macao Foundation.
D.~Li is also supported by the Specialized Research Fund for State
Key Laboratory of Solar Activity and Space Weather. CBSm consists of
four subsystems is operated by SDU. DART is an interferometric
imaging telescope supported by Chinese initiative Meridian Project
Phase 2. We thank the teams of DART, CBSm, MUSER, MSS-1, ASO-S,
GOES, and SDO, for their open data use policy. We acknowledged the
use of data from the Chinese MeridianProject.


\clearpage
\begin{figure}
\centering
\includegraphics[width=\linewidth,clip=]{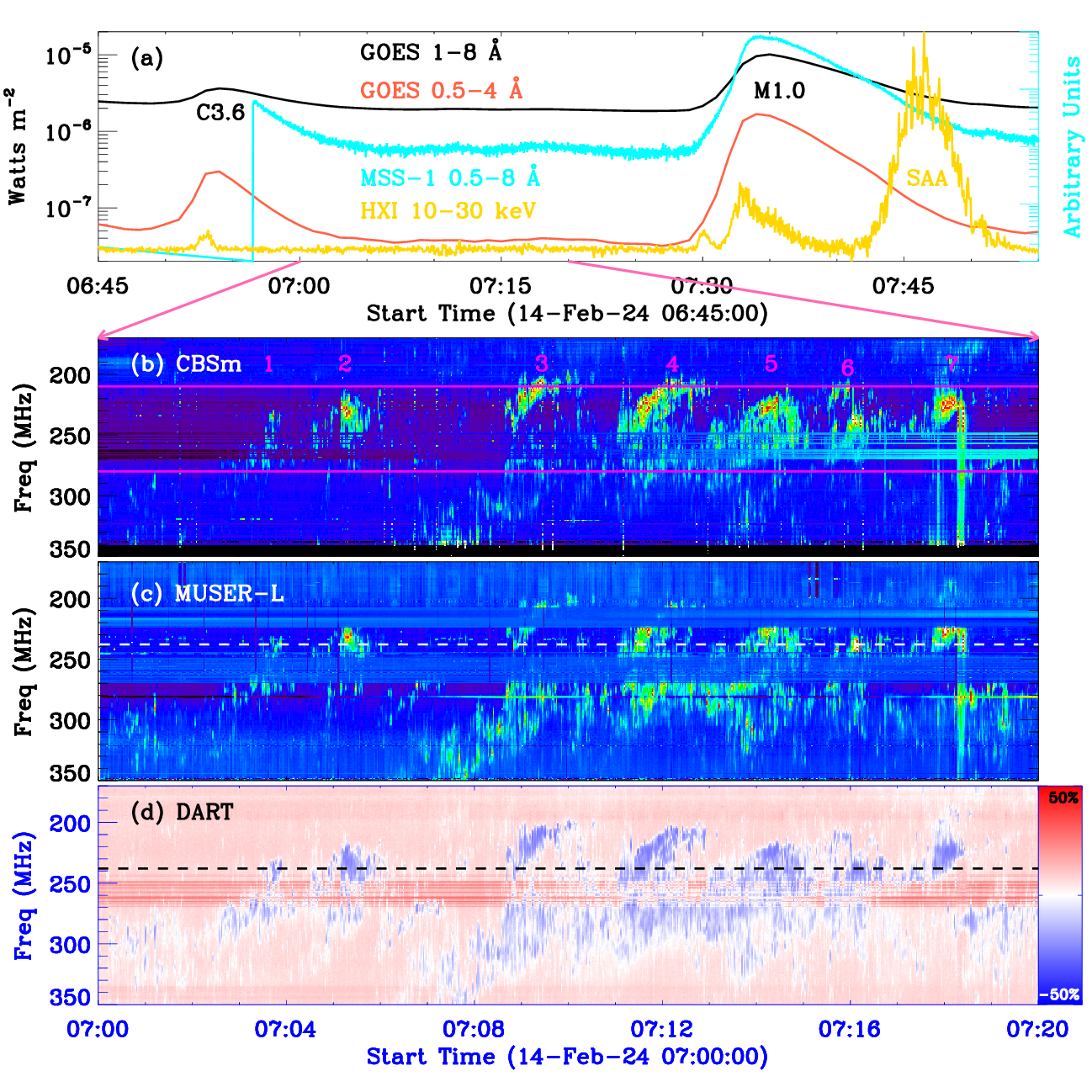}
\caption{Overview of type-I burst chains on 2024 February 14. (a)
SXR/HXR light curves recorded by GOES, MSS-1 and HXI during
06:45--07:55~UT, two solar flares are marked. (b) Dynamic spectrum
of CBSm between 07:00--07:20~UT. Two horizontal lines outline the
frequency range of solar metrewave bursts (1-7). (c)
Dynamic spectrum of MUSER-L. The white dashed line marks the
frequency at 238~MHz. (d) Dynamic spectrum of polarization measured
by DART. The dashed line marks the frequency at 238~MHz.
\label{over}}
\end{figure}

\begin{figure}
\centering
\includegraphics[width=\linewidth,clip=]{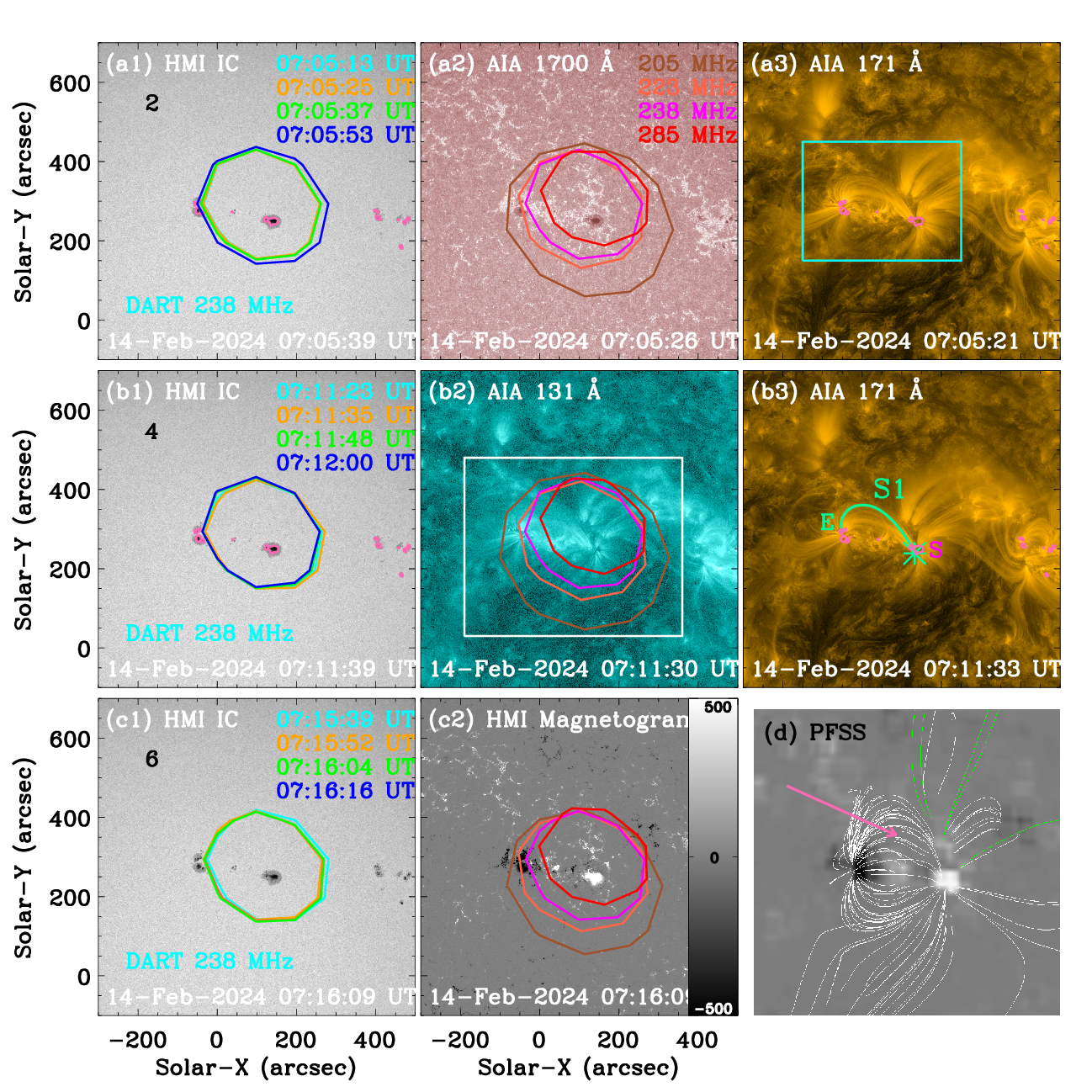}
\caption{Images of the source during type-I burst chains. (a1)-(c2)
Sub-maps with a FOV of about 800\arcsec~$\times$~800\arcsec~during
the burst chains~2, 4 and 6. They were measured by HMI continuum and
LOS magnetogram, AIA~1700~{\AA}, 171~{\AA}, and 131~{\AA}. The
overlaid contours are detected by DART, and the levels are set at
50\% of the local maximum values. The hot pink contours outline
sunspot umbrae. The cyan and white rectangles outline integrated
regions of AIA/HMI and DART. The green curves (S1) outline a loop
system generated for TD map, the asterisk indicates its start point.
(d) Magnetic field lines extrapolated from a PFSS model.
\label{img}}
\end{figure}

\begin{figure}
\centering
\includegraphics[width=\linewidth,clip=]{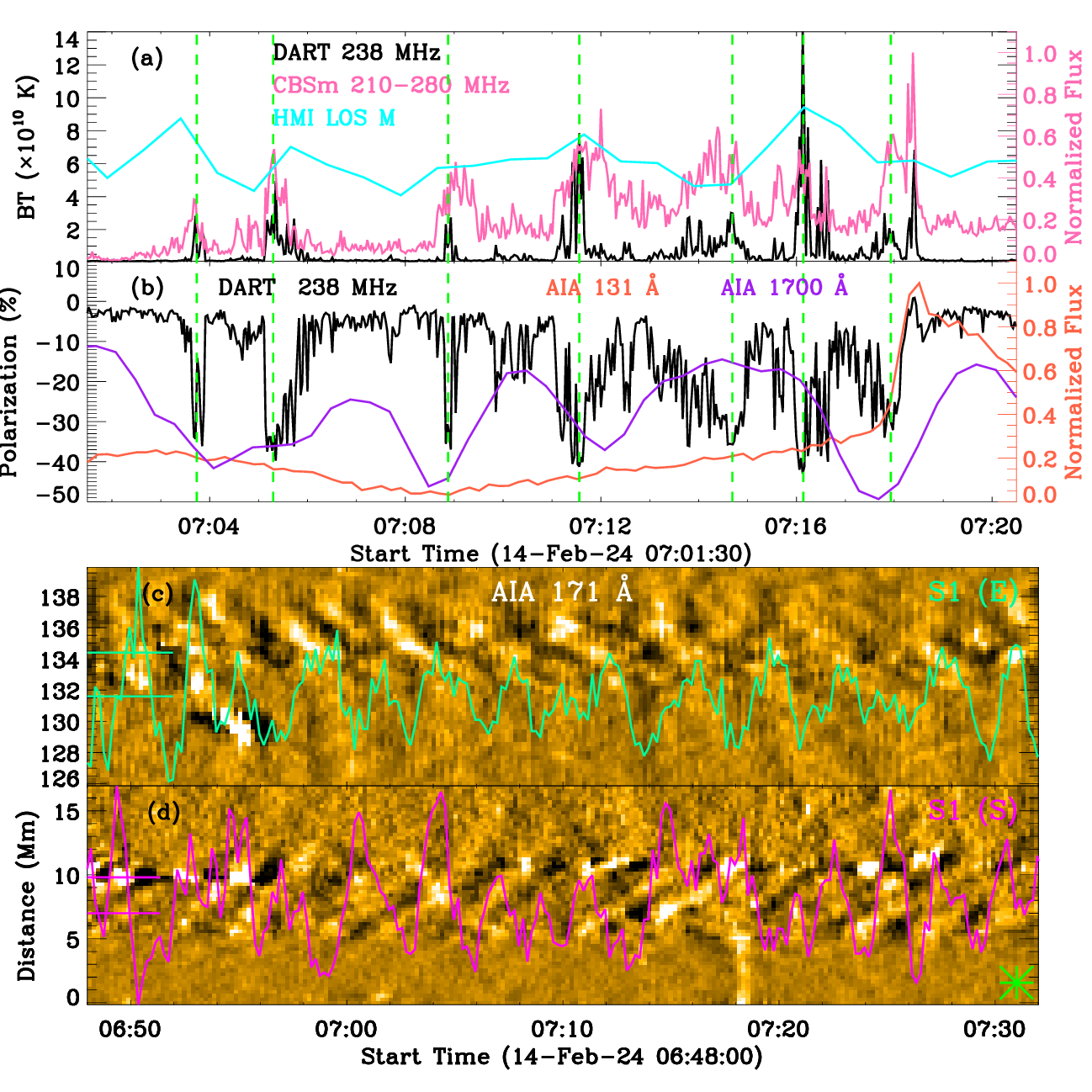}
\caption{(a) Local radio flux integrated over the source region at
DART 238~MHz, and full-disk radio flux integrated over the frequency
range of CBSm 210-280~MHz, as well as the net magnetic flux measured
by HMI. (b) Degree of circular polarization in the frequency of
DART~238~MHz. Local UV/EUV fluxes observed by AIA~1700~{\AA} and
131~{\AA}. (c)--(d) Detrended TD maps along S1 at AIA~171~{\AA}. The
overplotted curves are the time series extracted between two short
lines on the left. The asterisk indicates the start position.
\label{flux}}
\end{figure}

\begin{figure}
\centering
\includegraphics[width=\linewidth,clip=]{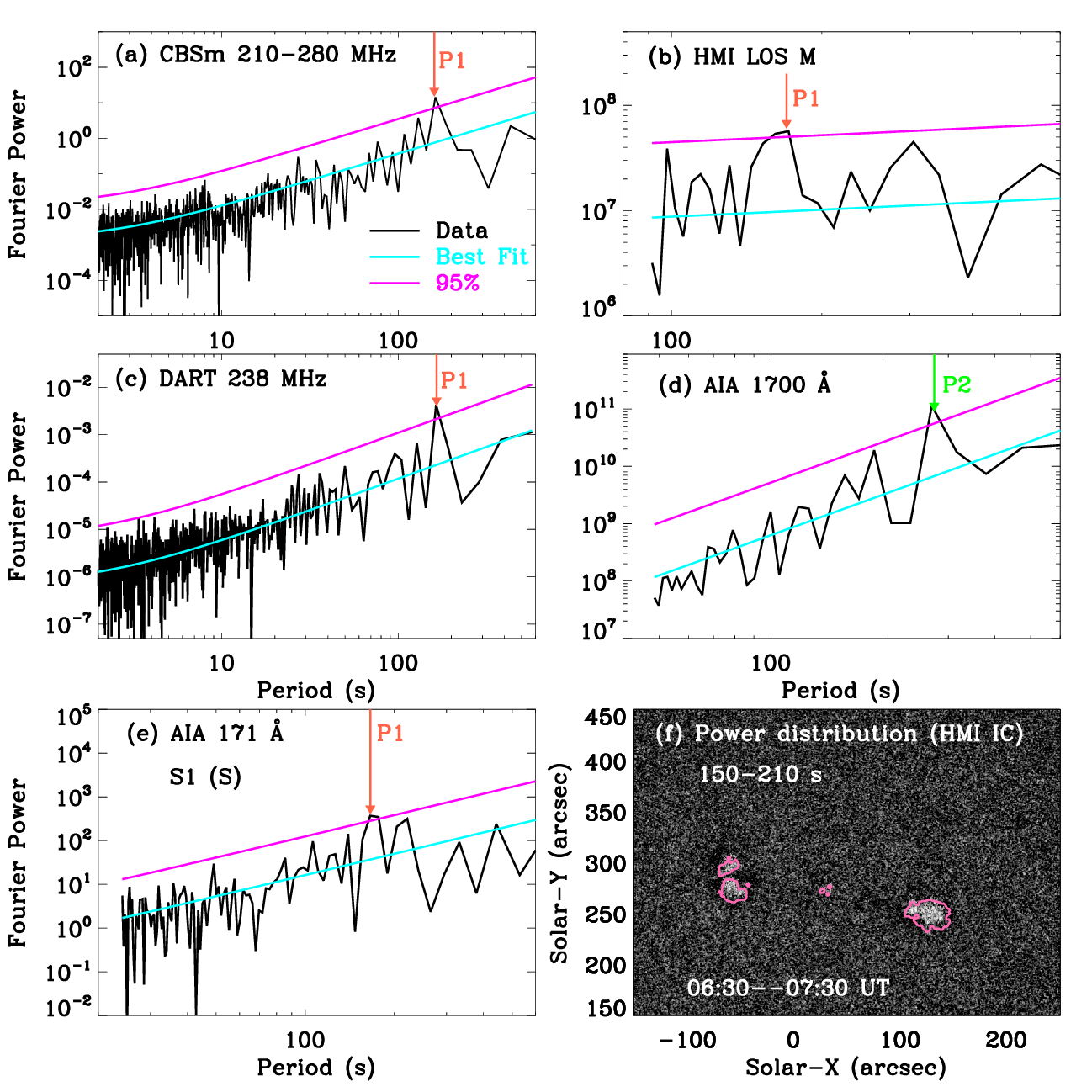}
\caption{Period analysis. (a)-(e) Fourier power spectra in wavebands
of CBSm~210-280~MHz, HMI LOS magnetogram, DART~238~MHz
(polarization), AIA~171~{\AA}, and 1700~{\AA}. The lines in each
panel represent the best fit (cyan) and the confidence level at 95\%
(magenta). The arrows outline the dominant periods above the
confidence level. (f) Fourier power maps averaged over 150-210~s,
obtained from HMI continuum maps. The overlaid contours outline the
sunspot umbrae. \label{per}}
\end{figure}

\begin{figure}
\centering
\includegraphics[width=\linewidth,clip=]{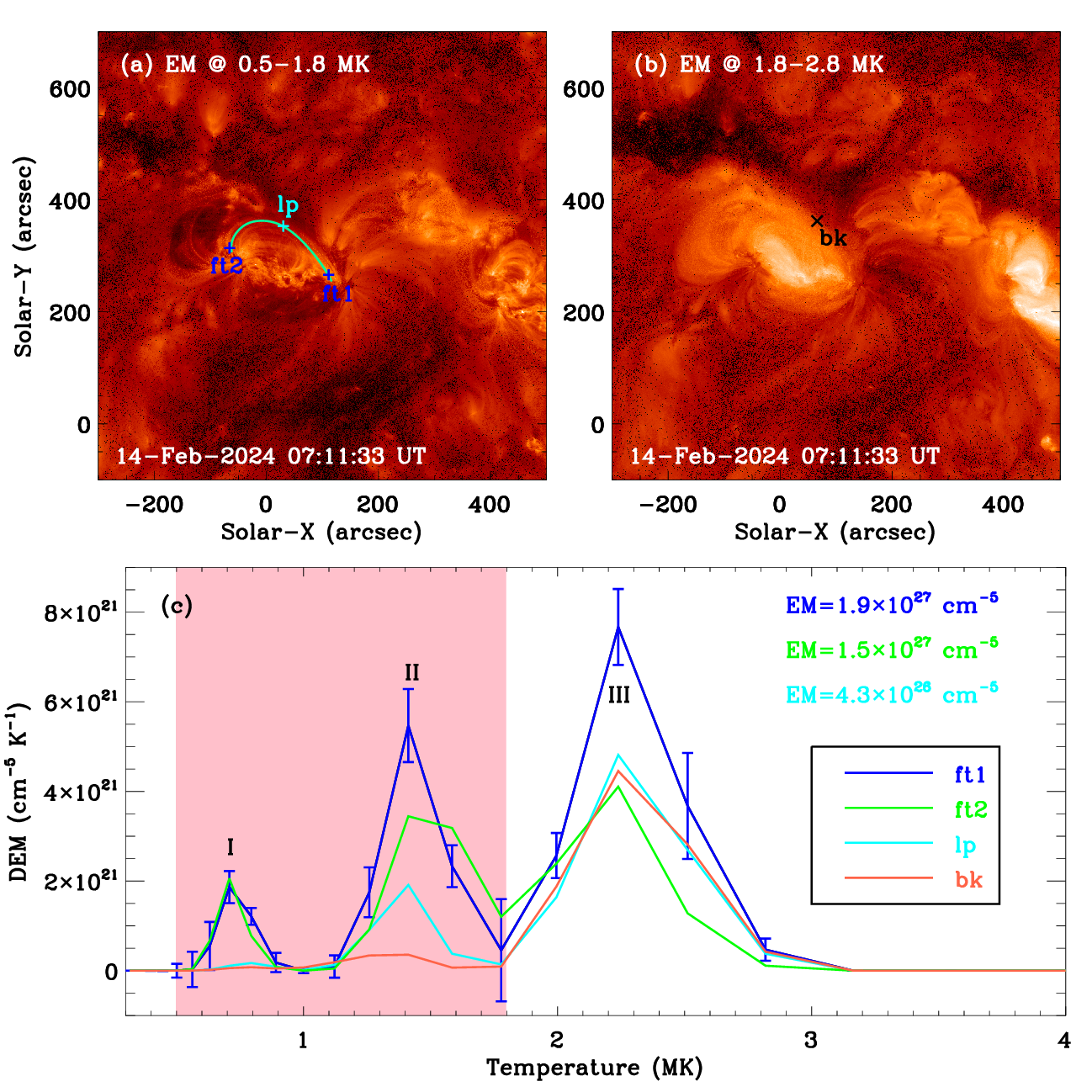}
\caption{DEM analysis results. (a) \& (b): EM maps with a FOV of
about 800\arcsec~$\times$~800\arcsec~integrated in temperature
ranges of 0.5-1.8~MK and 1.8-2.8~MK, respectively. The blue and cyan
pluses mark double footpoints (ft1 and ft2) and loop top (lp)
connected by the coronal loop (green curve), and the black cross
mark a background position. (c): DEM profiles at the selected
positions along the loop and in one location away from the loop. For
clarity, only the error bars at one position are shown. The pink
region outlines the temperature range used to integrate the EM.
\label{dem}}
\end{figure}

\clearpage
\begin{appendix}
\setcounter{figure}{0} \makeatletter
\renewcommand{\thefigure}{A\@arabic\c@figure}

\section*{Appendix~A}
Figure~\ref{add1} shows EUV/UV maps with a FOV of
$\sim$400$^{\prime\prime}$$\times$400$^{\prime\prime}$ during two
solar flares, which were measured by SDO/AIA in wavelengths of
131~{\AA} and 1700~{\AA}. The C3.6 peaked at about 06:54~UT and the
M1.0 flare reached its SXR maximum at about 07:35~UT. The HXR
radiation was reconstructed from the ASO-S/HXI data with a
HXI\_CLEAN algorithm, and the pixel scale was chosen to be 4\arcsec.
We note that the two flares located in the active region NOAA 13576
in the southern hemisphere.

Figure~\ref{add2} presents the supplemental material corresponding
to the animation of anim.mp4, including a radio dynamic spectrum and
multi-wavelength images. These same images appear in some select
panels of Figure~\ref{over} and Figure~\ref{img}. The animation
shows the temporal and spatial evolution of solar metrewave bursts.

\begin{figure}
\centering
\includegraphics[width=\linewidth,clip=]{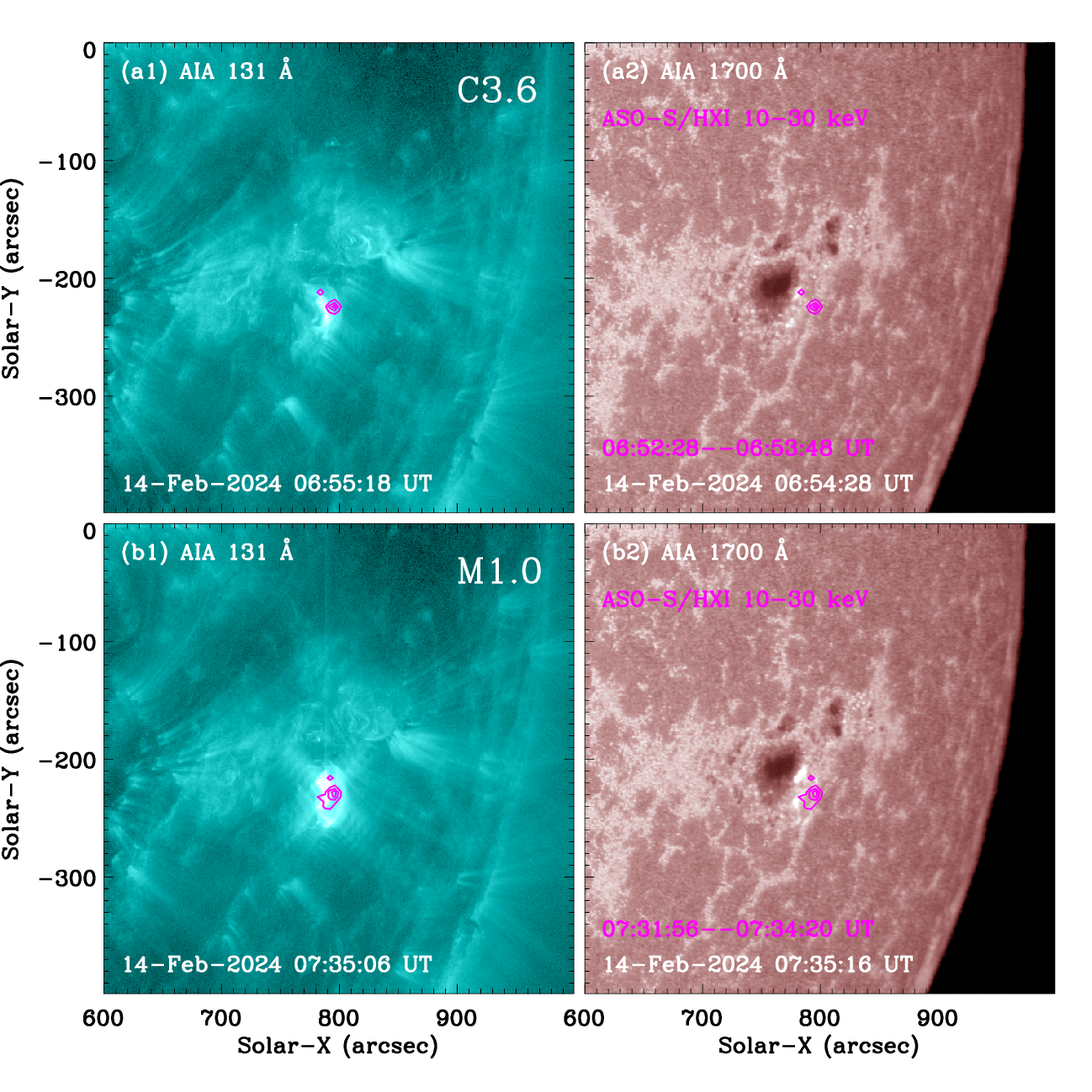}
\caption{Snapshots with a FOV of
400$^{\prime\prime}$$\times$400$^{\prime\prime}$ during two solar
flares in wavelengths of AIA~131~{\AA} and 1700~{\AA}. The overlaid
contours represent the HXR radiation observed by ASO-S/HXI.
\label{add1}}
\end{figure}

\begin{figure}
\centering
\includegraphics[width=\linewidth,clip=]{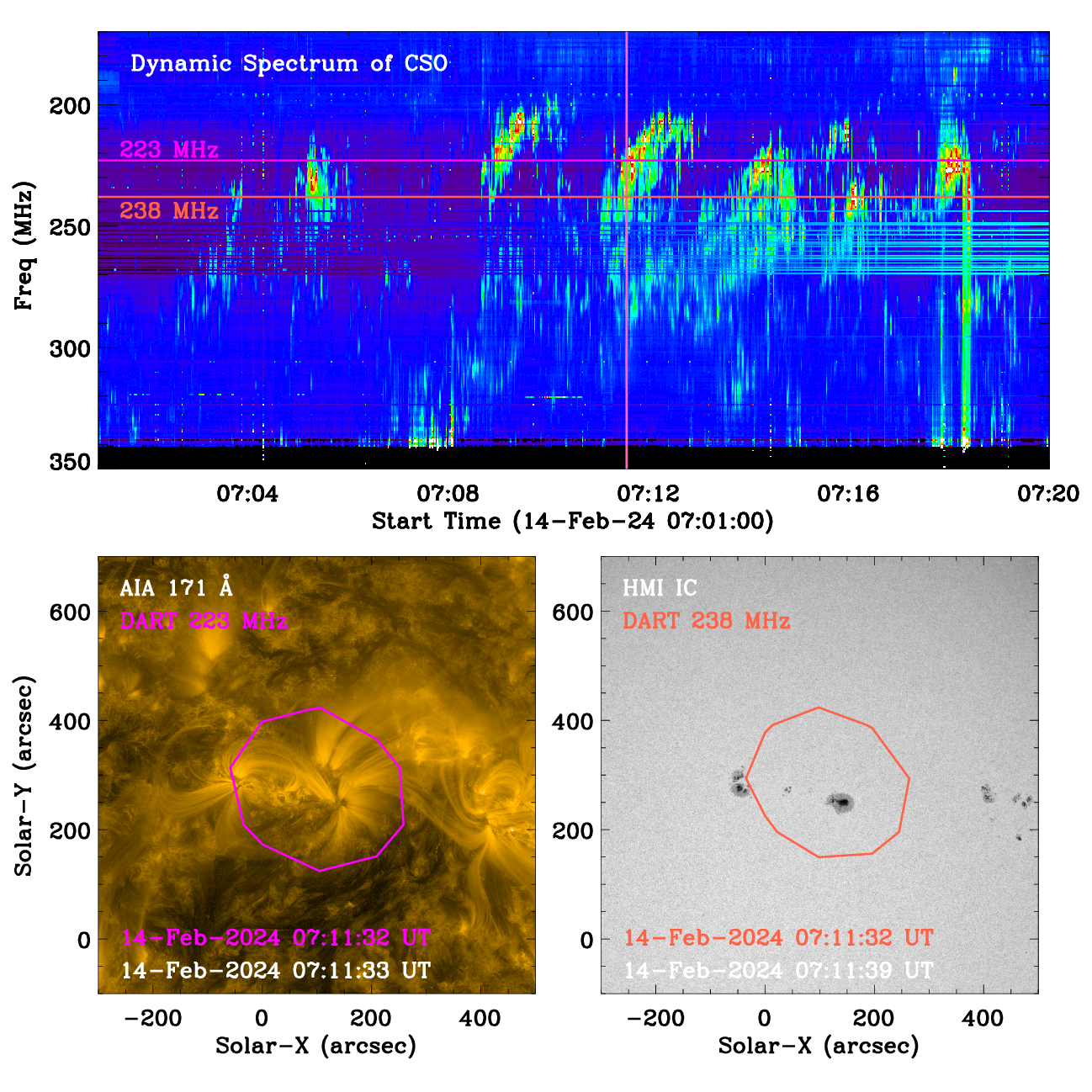}
\caption{Upper: Radio dynamic spectrum measured by CBSm. The
vertical line mark the time showed in the bottom panel, and two
horizontal lines outline the frequency at 223~MHz and 238~MHz.
Bottom: Sub-maps with a FOV of about 800\arcsec~$\times$~800\arcsec,
which were measured by AIA~171~{\AA} and HMI continuum, the overlaid
contours represent radio emissions in frequencies of 223~MHz and
238~MHz. The figure is associated with the online animation.
\label{add2}}
\end{figure}

\end{appendix}

\end{document}